\documentclass[%
reprint,
superscriptaddress,
 amsmath,amssymb,
 aps,
pra,
]{revtex4-1}

\usepackage{graphicx}
\usepackage{dcolumn}
\usepackage{bm}
\usepackage{physics}
\usepackage{siunitx}
\usepackage{mathtools}
\usepackage{extarrows}
\usepackage{bbold}

\usepackage{dsfont}

\usepackage{xr-hyper}
\usepackage{hyperref}
\hypersetup{
    colorlinks = true,
    allcolors = {blue},
    urlcolor = {black}
}
\usepackage[capitalize]{cleveref}
\usepackage{xfrac}

\usepackage{xcolor}
\newcommand{\AB}[1]{\textcolor{blue}{#1}}

\usepackage{soul}

\NewDocumentCommand{\evalat}{sO{\big}mm}{%
  \IfBooleanTF{#1}
   {\mleft. #3 \mright|_{#4}}
   {#3#2|_{#4}}%
}

\newcommand{\appropto}{\mathrel{\vcenter{
  \offinterlineskip\halign{\hfil$##$\cr
    \propto\cr\noalign{\kern2pt}\sim\cr\noalign{\kern-2pt}}}}}

\begin{document}

\title{Estimating the concentration of chiral media with bright squeezed light}

\author{Alexandre Belsley}
\email{alex.belsley@bristol.ac.uk}
\affiliation{%
Quantum Engineering Technology Labs, H. H. Wills Physics
Laboratory and Department of Electrical \& Electronic Engineering, University of Bristol, Bristol BS8 1FD, United Kingdom
}%
\affiliation{%
Quantum Engineering Centre for Doctoral Training, H. H. Wills Physics Laboratory and Department of Electrical \& Electronic Engineering, University of Bristol, Bristol BS8 1FD, United Kingdom
}%
\author{Jonathan C. F. Matthews}
\email{jonathan.matthews@bristol.ac.uk}
\affiliation{%
Quantum Engineering Technology Labs, H. H. Wills Physics
Laboratory and Department of Electrical \& Electronic Engineering, University of Bristol, Bristol BS8 1FD, United Kingdom
}%

\date{\today}

\begin{abstract}
The concentration of a chiral solution is a key parameter in many scientific fields and industrial processes. This parameter can be estimated to high precision by exploiting circular birefringence or circular dichroism present in optically active media. Using the quantum Fisher information formalism, we quantify the performance of Gaussian probes in estimating the concentration of chiral analytes. We find that bright-polarization squeezed state probes provide a quantum advantage over equally bright classical strategies that scales exponentially with the squeezing factor for a circularly birefringent sample. Four-fold precision enhancement is achievable using state-of-the-art squeezing levels and intensity measurements.
\end{abstract}
\maketitle
The precise estimation of physical quantities is ubiquitous in science and technology. Quantum resources can enhance measurement precision beyond that obtainable with classical strategies~\cite{Degen2017Jul, Polino2020Jun}. Notable applications include gravitational wave detection with squeezed light interferometry~\cite{Tse2019Dec, Acernese2019Dec}, sub-shot noise imaging~\cite{Brida2010Apr, Israel2014Mar,  Sabines-Chesterking2019Oct} and probing delicate samples~\cite{Wolfgramm2013Jan,Casacio2021Jun}.

The polarization degree of freedom has been the resource of choice in seminal quantum optics experiments such as the violation of Bell's inequalities~\cite{Aspect1982Jul}, quantum key distribution~\cite{Bennett1992Jan}, teleportation~\cite{Bouwmeester1997Dec} and super-resolving phase measurements using N00N states~\cite{Mitchell2004May}.
Besides being a useful degree of freedom for encoding quantum information, polarization can also be employed to probe physical properties of matter. When traversing optically active media, left (LCP) and right circularly polarized (RCP) light accumulate different phases --- circular birefringence --- and can undergo differential absorption --- circular dichroism. Practical use cases comprise the characterization of protein structures~\cite{Johnson1990Jan, Kelly2005Aug, Greenfield2006Dec}, nucleic acids~\cite{Kypr2009Apr}, and the conformation of biomolecules~\cite{Greenfield1969Oct}.

Characterizing optical phase and loss has been the object of many quantum metrology studies~\cite{Higgins2007Nov, Afek2010May, Zhou2015June, Birchall2020Apr, Gianani2021Dec, Belsley2022Jun}. Both phase and loss are sensitive to concentration, which is a key control parameter in several industrial processes such as pharmaceutical screening, material and food processing~\cite{Brady2008Apr}. In these applications, chiral properties of matter are often leveraged to probe dilute analytes~\cite{Djerassi1961Sep,Nguyen2006Jun,Ranjbar2009Aug}.

Recently, theoretical studies have investigated the use of quantum probes to characterize the net phase and absorption coefficients of chiral media~\cite{Rudnicki2020Aug, Ioannou2021Nov, Wang2021Dec}. Proof of principle experiments using polarization entangled photon pairs~\cite{Tischler2016Oct} and heralded single photons~\cite{Yoon2020Jun} performed sub-shot noise estimation of the optical rotatory dispersion and concentration of a sucrose solution respectively.

Here, we extend the above studies to investigate estimating the concentration of optically active media, leveraging circular birefringence or circular dichroism. Using quantum estimation theory, we determine the highest precision achievable with bright Gaussian state probes. We find that polarization squeezed state probes can provide a four-fold precision enhancement over equally bright classical strategies in a low loss, circularly birefringent system. The quantum advantage is found to scale exponentially with the squeezing factor in this scenario. We also predict an order of magnitude precision enhancement for highly dilute circularly dichroic solutions.  

\begin{figure}[t!]
    \centering
    \includegraphics[width=0.8\linewidth]{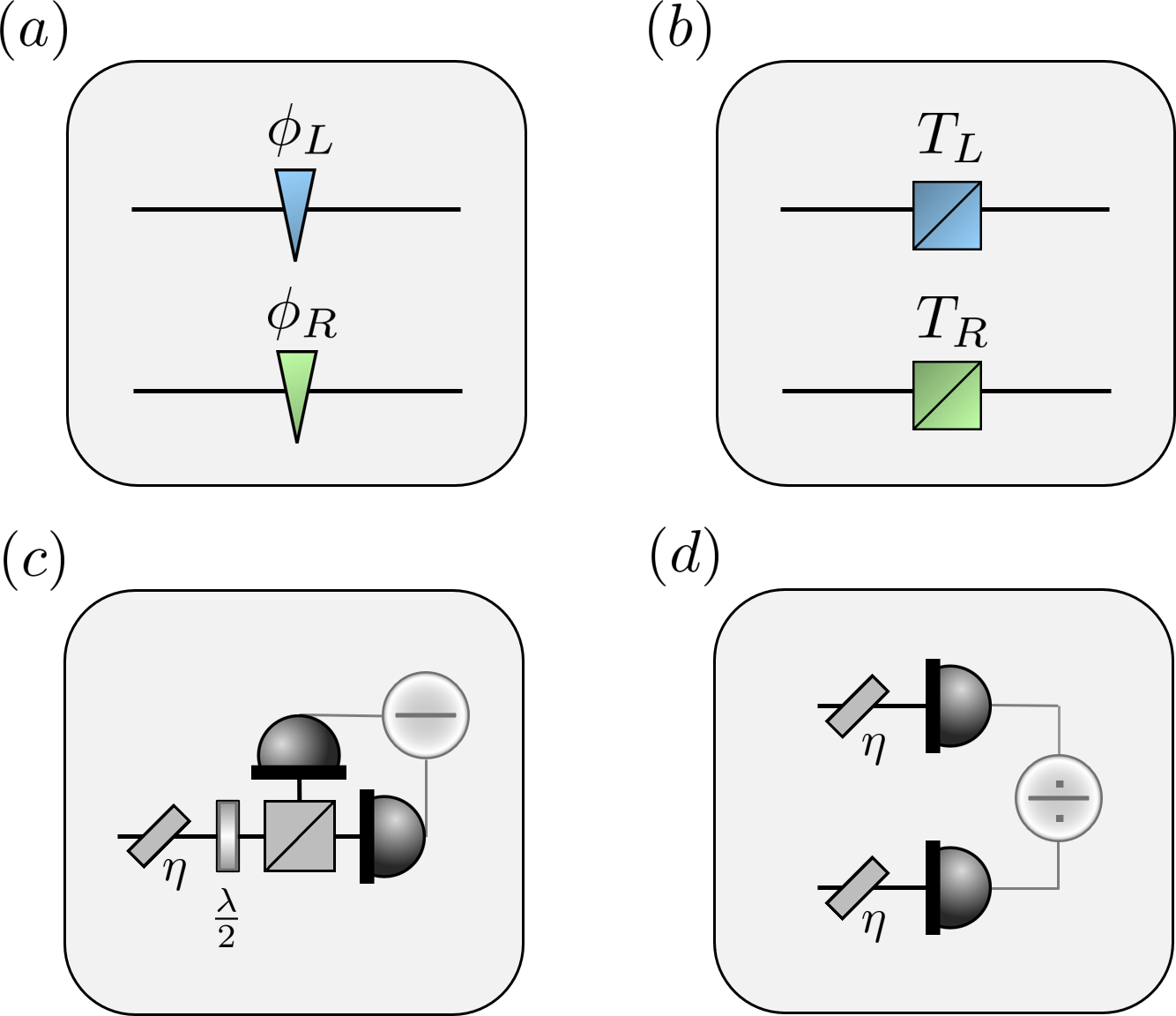}
    \caption{Quantum-mechanical models of (a) circular birefringece and (b) circular dichroism where the LCP and RCP modes undergo phase shifts $\phi_{L/R}$ and attenuation $T_{L/R}$, respectively. Optimal detection schemes for (c) two-mode polarization squeezed state and (d) twin single-mode amplitude squeezed state probes. External losses are modeled using a beamsplitter with transmission $\eta$ before each detector and $\lambda/2$ denotes a half-waveplate.}
    \label{fig:Main_sketch}
\end{figure}

\newpage
The concentration $C$ can be estimated with a precision bounded by~\cite{Helstrom1969Jun}
\begin{equation}
    \Delta^2 \tilde{C} \overset{1}{\geq} \frac{1}{\nu\, \mathcal{F}(C)} \overset{2}{\geq} \frac{1}{\nu\, \mathcal{Q}(C)} \,.
\label{eq:QCRB_CRB}
\end{equation}
\noindent The Cram\'er-Rao bound (inequality 1) relates the variance of the unbiased estimator $\tilde{C}$ to the classical Fisher information, $\mathcal{F}(C)$, for a given strategy with $\nu$ trials. Maximizing $\mathcal{F}(C)$ over all physical measurements leads to the quantum Fisher information (QFI), $\mathcal{Q}(C)$~\cite{Braunstein1994May}. Inequality 2 in \cref{eq:QCRB_CRB}, the Quantum Cram\'er-Rao bound, specifies the best precision achievable in estimating $C$ for a given channel and probe state. In the following, we assume that all other system parameters are known.

For many systems of interest, circular birefringence is reasonably strong in the visible and near infrared where quantum probes are more readily available. In these spectral regions, circular dichroism is generally vanishingly small. Hence, we first concentrate on estimating concentration from circular birefringence alone and neglect circular dichroism.

In media with optical activity, the LCP and RCP eigenpolarizations of light experience unequal refractive indices. Upon propagating through a chiral medium of length $l$, these components accumulate a differential phase~\cite{Condon1937Oct}
\begin{equation}
    \Delta \phi = \phi_{R} - \phi_{L} = \delta \gamma\, C\,l \,, 
    \label{eq:circular_birefringence}
\end{equation}
\noindent where $\delta \gamma$ is the optical rotatory power (see \cref{fig:Main_sketch} (a)). When linearly polarized light is incident on a circularly birefringent medium, the different phases acquired by the LCP and RCP eigenmodes result in a net rotation of the polarization angle upon exit. The angle of rotation is proportional to the concentration $C$ of the analyte, leading to a straightforward estimate of $C$ provided the optical rotatory power per path length $\delta \gamma/l$ is known. 

Because we are interested in determining $C$ through the rotation angle that linearly polarized light undergoes upon traversing a circularly birefringent medium, it is opportune to consider Gaussian states that have reduced fluctuations in the equatorial plane of the Poincare sphere. We thus consider the following single spatial mode, two polarization mode squeezed state
\begin{equation}
\ket{\psi_G} =\hat{D}_H(\alpha)\,\hat{S}_H(s,\theta)\, \hat{S}_V(s,\theta)\,\ket{0_H\,, 0_V}\,,
\label{eq:twin_bSMSS_probe_state}
\end{equation}
\noindent where $\hat{S}_i(s,\theta) = e^{s (e^{-i \theta}\, \hat{a}_i^2 - e^{i \theta}\, \hat{a}_i^{\dag 2})/2}$ and $\hat{D}_i(\alpha_i) = e^{\alpha_i(\hat{a}_i^\dagger - \hat{a}_i)}$ are the single-mode squeezing and displacement operators, respectively. The subscripts $H$ and $V$ denote the horizontal and vertical polarization modes. This state is fully characterized by a displacement vector $\boldsymbol{d}$ with elements $d_i = \langle \hat{A}_i \rangle$ and a covariance matrix $\boldsymbol{\Sigma}$ with elements $\Sigma_{ij} = \langle \hat{A}_i\, {\hat{A}_j}^\dagger + {\hat{A}_j}^\dagger\, \hat{A}_i \rangle - 2 \langle \hat{A}_i \rangle \langle {\hat{A}_j}^\dagger \rangle$ where $\hat{A} = \begin{pmatrix} \hat{a}_H, \hat{a}_V, {\hat{a}_H}^\dagger,{\hat{a}_V}^\dagger \end{pmatrix}^\intercal$~\cite{Weedbrook2012May}.

After traversing a circularly birefringent medium, the displacement vector becomes $\boldsymbol{\tilde{d}} = \begin{pmatrix} 
d_H,  & d_V, & d_H^*, & d_V^*
\end{pmatrix}^\intercal$ with $d_H = \alpha \cos(\Delta\phi/2)\,e^{i(\phi_L+\phi_R)/2}$ and $d_V = \alpha \sin(\Delta\phi/2)\,e^{i(\phi_L+\phi_R)/2}$. The 4x4 covariance matrix $\boldsymbol{\tilde{\Sigma}}$ has non-zero diagonal elements  $\Sigma_{ii} = \cosh(2s)$ and $\Sigma_{13} = \Sigma_{24} = -\sinh(2s) e^{i(\theta+\phi_L+\phi_R)/2}= \Sigma_{31}^* = \Sigma_{42}^*$.

The QFI for the Gaussian probe state in \cref{eq:twin_bSMSS_probe_state} is given by (see Supplementary Materials A)
\begin{align}
     \mathcal{Q}_G (C) = &4 l^2\, \delta \gamma^2 \sinh(2s)^2 \nonumber \\ &+ \abs{\alpha}^2\, l^2\, \delta \gamma^2 \left(\cosh2s+\abs{\sin\Delta\phi}\,\sinh2s\right)\,,
\label{eq:QFI_bTSMSS}
\end{align}
\noindent at the optimal squeezing angle $\theta = \pi - \phi_L - \phi_R$. The second term in \cref{eq:QFI_bTSMSS} represents the information contributed by the displacement vector $2 (\partial_C\, \boldsymbol{\tilde{d}})^\intercal\, \boldsymbol{\tilde{\Sigma}}^{-1}\, (\partial_C\, \boldsymbol{\tilde{d}})$~\cite{Safranek2015Jul}, which we call the bright term. When $\abs{\alpha}^2 \gg 4 \sinh{2s} \tanh{2s}$, the bright term dominates over the first term, i.e. the vacuum contribution. Even for squeezing levels as large as $s=1.8$, a value of $\abs{\alpha}^2 = 750$  will ensure that the bright term is more than ten times greater than the vacuum term. As such, the bright term provides a lower bound on the QFI
\begin{equation}
    \mathcal{Q}_G (C) \geq \abs{\alpha}^2\, l^2\, \delta \gamma^2 \left(\cosh2s+\abs{\sin\Delta\phi}\,\sinh2s\right)\,,
\label{eq:QFI_bTSMSS_lb}
\end{equation}
which, in practice, accounts for the near entirety of the QFI.

The QFI for the coherent-state probe $\ket{\alpha_H,0_V}$ is readily obtained by setting the squeezing factor $s \rightarrow 0$, yielding
\begin{equation}
    \mathcal{Q}_{\text{SQL}}(C) = \abs{\alpha}^2\, l^2\, \delta \gamma^2\,,
\label{eq:QFI_coherent}    
\end{equation}
\noindent which corresponds to the standard quantum limit (SQL). 

\begin{figure}
    \centering
    \includegraphics{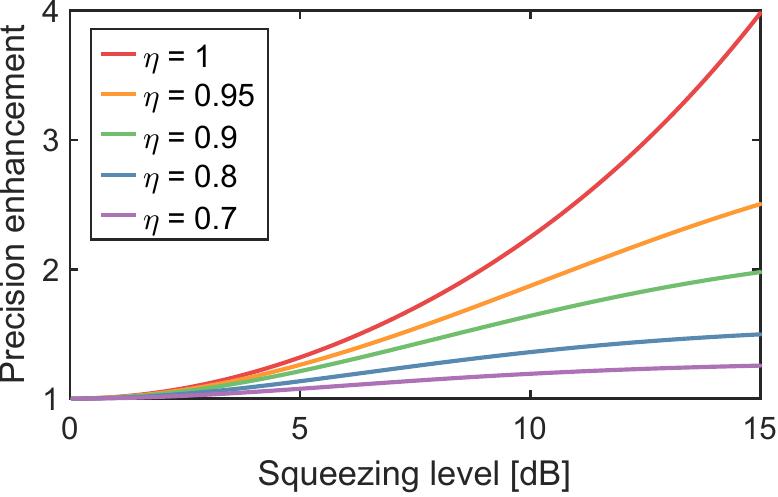}
    \caption{Precision enhancement in estimating the concentration of a chiral analyte via circular birefringence. The enhancement beyond the SQL is plotted as a function of the squeezing level for the two-mode polarization squeezed state $\ket{\psi_G}$ and different system efficiencies $\eta$. In low loss systems, state-of-the-art squeezing levels yield up to a fourfold precision enhancement over an equally bright classical strategy.}
    \label{fig:QFI_circular birefringence}
\end{figure}

Comparing \cref{eq:QFI_bTSMSS_lb,eq:QFI_coherent}, the two-mode polarization squeezed probe $\ket{\psi_G}$ surpasses the SQL for any finite level of squeezing, providing a quantum advantage $\mathcal{Q}_G / \mathcal{Q}_{\text{SQL}} \gtrsim e^{2s}$ in the limit of large squeezing. The current squeezing level record is 15 dB ($s \approx 1.73$)~\cite{Vahlbruch2016Sep} which, for low loss systems, translates to a fourfold precision enhancement beyond the SQL in estimating the concentration of a chiral analyte as shown in~\cref{fig:QFI_circular birefringence}.

As an example, we consider estimating the concentration of a dilute $1 \%$ w/w aqueous sucrose solution where $\delta \gamma = \SI{1.16}{ \centi\meter\cubed \per  \gram \per \deci\meter}$ at $\lambda = \SI{589}{\nano\meter}$~\cite{Lowry1924Jan,Harris2002Jul}. For a standard cuvette length $l = \SI{1}{\centi\meter}$ and $\abs{\alpha}^2 = 10^9$, we find that the two-mode polarization squeezed state $\ket{\psi_G}$ with a squeezing level $s=1$ ($\approx \SI{8.7}{\dB}$) can attain a relative precision $\Delta C /C = 0.008$, twice better than achievable with a coherent-state probe $\Delta C/C = 0.016$.

Having derived the fundamental precision limit attainable with the two-mode polarization squeezed state $\ket{\psi_G}$ given in \cref{eq:QFI_bTSMSS}, we now show that a balanced detection scheme can saturate it.

The optimal measurement operator is $\hat{S} = ({\hat{a}_H}^\dagger \cos{\xi} + {\hat{a}_V}^\dagger \sin{\xi}) \cdot \text{c.c.} - ({\hat{a}_H}^\dagger \sin{\xi} - {\hat{a}_V}^\dagger \cos{\xi}) \cdot \text{c.c.}$, where $\xi$ is a rotation angle and $\text{c.c.}$ denotes the complex conjugate. Experimentally, $\hat{S}$ can be determined using a half-waveplate inducing a rotation $\xi$ followed by a polarization beamsplitter to separate the $H$ and $V$ components, whose intensities are measured by separate detectors and then subtracted as depicted in \cref{fig:Main_sketch} \AB{(c)}.

This measurement strategy yields a mean value $\langle \hat{S} \rangle =\abs{\alpha}^2  \cos (\Delta \phi-2 \xi)$ and variance $\langle \Delta^2 \hat{S} \rangle = \abs{\alpha}^2 (\cosh{2s}-\cos{\theta} \sinh{2s})$ in the bright limit where $\abs{\alpha}^2 \gg \sinh^2 2s$.

For Gaussian probes, we can readily obtain the variance in estimating the concentration using error propagation
\begin{equation}
    \Delta^2 C = \langle \Delta^2 \hat{S} \rangle \abs{\frac{\partial \langle \hat{S} \rangle}{\partial C}}^{-2} = \left(\abs{\alpha}\,l\, \delta\gamma\,e^{s}\right)^{-2}\,,
\label{eq:variance_S}
\end{equation}
\noindent for a rotation angle $\xi = (2 \Delta \phi - \pi)/4$ induced by the waveplate. For small dilute samples where $\abs{\sin \Delta \phi} \approx 0$, $\Delta^2 C$ is equal to the reciprocal of \cref{eq:QFI_bTSMSS} such that a measurement of the operator $\hat{S}$ saturates the QCRB.

Note that this differs from the conventional measurement carried out for a classical probe state, where one would typically measure the circular birefringence induced rotation angle by crossing a linear polarizer with the transmitted beam. Practically, the conventional strategy is susceptible to DC noise and intensity fluctuations of the incident probe field while the balanced detection scheme advocated is immune to common mode noise sources.

External system losses such as non-unitary channel transmission and photodetection efficiency are modeled using a fictitious beamsplitter with transmittance $\eta$ as shown in \cref{fig:Main_sketch} \AB{(c)} and \AB{(d)}. An expression for the QFI considering external system losses is derived in Supplementary Materials A.

Apart from circular birefringence, optically active media can also exhibit circular dichroism. This phenomenon refers to the differential absorption experienced by LCP and RCP light traversing a chiral medium with length $l$. According to the Beer–Lambert–Bouguer law, the absorbance of each circularly polarized component is directly proportional to the concentration, i.e.~\cite{Woody1995}
\begin{equation}
    A_{i} = \epsilon_{i}\, C\, l\,,
\end{equation}
\noindent where $\epsilon_i$ is the molar extinction coefficient.
This effect is typically strongest in the UV region where bright quantum states are challenging to generate. For completeness, we quantify the performance of bright Gaussian probes in estimating the concentration $C$ of a circularly dichroic analyte.

The differential absorption of the LCP and RCP modes can be modeled quantum mechanically using fictitious beamsplitters with intensity transmissions $T_i = 10^{-A_i}$ as depicted in \cref{fig:Main_sketch} \AB{(b)}. Ref.~\cite{Ioannou2021Nov} used this model to estimate the transmission circular dichroism parameter $T_L-T_R$. 
However, to estimate the concentration $C$, one should instead resort to absorbance circular dichroism where the parameter of interest is the difference in absorbance $\Delta A = A_L - A_R$ between the LCP and RCP modes. For most biological samples, this quantity is of the order $3 \cdot 10^{-4}$~\cite{Kelly2005Aug}. A practical detection strategy is to measure the intensities $I_i$ of the LCP and RCP modes after traversing the chiral medium as shown in \cref{fig:Main_sketch} \AB{(d)}. From the ratio $I_L/I_R = 10^{\Delta A}$, one can readily obtain an estimate of the concentration $C$ provided the molar circular dichroism per path length $\Delta \epsilon/l$ is known.

Fock or squeezed vacuum probes have been shown to be optimal in single-mode transmission estimation on a mean input photon basis~\cite{Adesso2009Apr, Nair2018Dec}. Single-mode bright squeezed states were shown to approach the performance of these probes for large squeezing values~\cite{Woodworth2020Nov}. Bright squeezed state probes can be generated with macroscopic numbers of photons unlike Fock or squeezed vacuum states, resulting in an overall higher precision estimate of the transmission as the QFI is proportional to the mean number of probe photons. Given that each circularly polarized input mode in our system is independent and undergoes a transmission $T_i$, the twin single-mode amplitude squeezed state
\begin{equation}
    \ket{\psi_B}=\hat{D}_L(\alpha)\hat{D}_R(\alpha) \hat{S}_L(s,\theta) \hat{S}_R(s,\theta) \ket{0_L, 0_R}\,,
    \label{eq:Probe_twin_SMASS}
\end{equation}
\noindent is a good candidate for estimating the concentration $C$. As derived in Supplementary Materials B, this probe yields an uncertainty
\begin{equation}
    \Delta^2 C = \beta \left[2(e^{-2s}-1) + 1/T_R + 1/T_L \right]\,,
    \label{eq:Concentration_uncertainty}
\end{equation}
\noindent where $\beta = (\abs{\alpha} \Delta \epsilon\,l\, \ln 10)^{-2}$. Compared to a coherent-state probe where $s=0$, the squeezed probe $\ket{\psi_B}$ provides a precision enhancement that scales approximately as $1/\sqrt{1-T_i}$. For dilute samples where $T_i \gtrsim 0.9$, the quantity $1/T_R + 1/T_L \lesssim 2.22$. In this case, a large squeezing factor $s$ results in a standard deviation $\Delta C \approx \sqrt{0.22 \beta}$, about three times smaller than that obtainable with a coherent-state probe. The larger $T_i$ is, the greater the quantum advantage provided by bright squeezing reaching an order of magnitude for $T_i \approx 0.99$.

\clearpage
In summary, we have quantified the performance of bright squeezed states in estimating the concentration of chiral solutions. 

For media exhibiting circular birefringence, we found that two-mode polarization squeezed state probes can provide a four-fold precision enhancement over an equally bright classical strategy in low loss systems with state-of-the art squeezing levels. The quantum advantage scales exponentially with the squeezing factor and balanced detection is the optimal measurement strategy for probing dilute analytes.

In the case of absorbance circular dichroism, twin single-mode amplitude squeezed states can outperform coherent-state probes. An order-of-magnitude higher precision estimate of the concentration is predicted for highly dilute samples. The development of bright squeezed states in the UV spectral region would allow these performance gains to be attained.

Our results benchmark the potential precision enhancement that Gaussian quantum probes with a macroscopic number of photons can provide in estimating the concentration of chiral media. They are especially pertinent in characterizing dilute chiral solutions.

\paragraph*{Acknowledgments ---} We thank Animesh Datta for helpful discussions. A.B. and J.C.F.M. acknowledge support from the ERC starting grant ERC-2018-STG 803665. A.B. also acknowledges support from the EPSRC grant EP/S023607/1.

\twocolumngrid
\bibliography{mainbib.bib}

\clearpage
\onecolumngrid
\begin{center}
\textbf{\large Supplementary Materials}
\end{center}

\setcounter{equation}{0}
\setcounter{figure}{0}
\setcounter{table}{0}
\makeatletter
\renewcommand{\theequation}{S\arabic{equation}}
\renewcommand{\thefigure}{S\arabic{figure}}
\renewcommand{\bibnumfmt}[1]{[#1]}
\renewcommand{\citenumfont}[1]{#1}

Here, we derive \cref{eq:QFI_bTSMSS,eq:Concentration_uncertainty} of the main text quantifying the performance of squeezed state probes in estimating the concentration $C$ of a medium that exhibits circular birefringence or circular dichroism.

\section*{A. Circular birefringence}
\label{sec:appendix_A}

In the presence of external system losses $\eta$, the displacement vector for the two-mode polarization squeezed state probe in \cref{eq:twin_bSMSS_probe_state} of the main text is given by
\begin{equation}
    \boldsymbol{\tilde{d}} = \sqrt{\eta} \begin{pmatrix} 
\alpha \cos(\Delta\phi/2)\,e^{i(\phi_L+\phi_R)/2}  \\ \alpha \sin(\Delta\phi/2)\,e^{i(\phi_L+\phi_R)/2} \\ \alpha^* \cos(\Delta\phi/2)\,e^{-i(\phi_L+\phi_R)/2} \\ \alpha^* \sin(\Delta\phi/2)\,e^{-i(\phi_L+\phi_R)/2}
\end{pmatrix}\,.
\end{equation}

\noindent The covariance matrix takes the following form
\begin{equation}
    \boldsymbol{\tilde{\Sigma}} = \begin{pmatrix}
1-\eta + \eta \cosh{2s} & 0 & -\eta \sinh(2s) e^{i(\theta+\phi_L+\phi_R)/2} & 0 \\
0 & 1-\eta + \eta \cosh{2s} & 0 & -\eta \sinh(2s) e^{i(\theta+\phi_L+\phi_R)/2} \\
-\eta \sinh(2s) e^{-i(\theta+\phi_L+\phi_R)/2} & 0 & 1-\eta + \eta \cosh{2s} & 0 \\
0 & -\eta \sinh(2s) e^{-i(\theta+\phi_L+\phi_R)/2} & 0 & 1-\eta + \eta \cosh{2s} \\
    \end{pmatrix}\,.
\end{equation}

The QFI for a two-mode Gaussian probe state is given by~\cite{Safranek2015Jul}
\begin{align}
    \mathcal{Q}_G(C) = &\frac{1}{2\left(\begin{vmatrix}\boldsymbol{\tilde{\sigma}}\end{vmatrix}-1\right)} \bigg\{\begin{vmatrix}\boldsymbol{\tilde{\sigma}}\end{vmatrix} \Tr[\left(\begin{vmatrix}\boldsymbol{\tilde{\sigma}}\end{vmatrix}^{-1} \partial_C \begin{vmatrix}\boldsymbol{\tilde{\sigma}}\end{vmatrix}\right)^2] + \sqrt{\abs{\mathbb{1}+\boldsymbol{\tilde{\sigma}}^2}} \Tr\left[\left(\left(\mathbb{1}+\boldsymbol{\tilde{\sigma}}^2 \right)^{-1} \partial_C \boldsymbol{\tilde{\sigma}} \right)\right]  \nonumber\\
    &+ 4(\lambda_1^2 -\lambda_2^2)\left(\frac{\partial_C\lambda_2^2}{\lambda_2^4-1}-\frac{\partial_C\lambda_1^2}{\lambda_1^4-1}\right)\bigg\} + 2 (\partial_C\, \boldsymbol{\tilde{d}})^\intercal\, \boldsymbol{\tilde{\sigma}}^{-1}\, (\partial_C\, \boldsymbol{\tilde{d}})\,,
\end{align}
\noindent where $\mathbb{1}$ denotes the identity matrix, $\boldsymbol{\tilde{\sigma}} = k \cdot \boldsymbol{\tilde{\Sigma}}$ with $k = \text{diag}(1,1,-1,-1)$ is the sympletic form of the covariance matrix and $\lambda_i$ are its sympletic eigenvalues. The first term in this expressions evaluates to
\begin{equation}
    \frac{4 \eta^2\, l^2\, \delta \gamma^2 \sinh(2s)^2}{1-\eta + \eta^2 + (1-\eta) \eta \cosh(2s)}\,,
\end{equation}
\noindent while the information contained in the displacement vector $2 (\partial_C\, \boldsymbol{\tilde{d}})^\intercal\, \boldsymbol{\tilde{\Sigma}}^{-1}\, (\partial_C\, \boldsymbol{\tilde{d}})$ is equal to
\begin{equation}
   \eta \abs{\alpha}^4\, l^2\, \delta \gamma^2 \left(1-\eta + \eta \cosh2s+\eta \abs{\sin\Delta\phi}\,\sinh2s\right) \,.
\end{equation}
\noindent In the absence of external system losses, i.e. $\eta=1$, the QFI reduces to \cref{eq:QFI_bTSMSS} in the main text, that is
\begin{equation}
     \mathcal{Q}_G (C) = 4 l^2\, \delta \gamma^2 \sinh(2s)^2 + \abs{\alpha}^2\, l^2\, \delta \gamma^2 \left(\cosh2s+\abs{\sin\Delta\phi}\,\sinh2s\right)\,.
\end{equation}

\section*{B. Circular dichroism}
\label{sec:appendix_B}

Consider the squeezed state probe $\ket{\psi_B}$ given in \cref{eq:Probe_twin_SMASS} and the measurement strategy outlined in the main text to estimate the concentration $C$ of a circularly dichroic medium. Briefly, $C$ can be determined from the ratio between the transmitted intensities $I_i$ of the LCP and RCP modes, i.e. $R = I_L/I_R = 10^{\Delta \epsilon C l}$.

The mean value and variance for an intensity measurement are $\langle \hat{n}_i \rangle = \langle \hat{a}^\dagger \hat{a} \rangle$ and $\Delta^2 \hat{n}_i = \langle \hat{n}_i^2 \rangle -\langle \hat{n}_i \rangle^2$, respectively. In the bright limit, these quantities are $\langle \hat{n}_i \rangle = T_i (\abs{\alpha}^2+\sinh^2 s)$ and $\Delta^2 \hat{n}_i = \abs{\alpha}^2 T_i (1+2 T_i \sinh^2 s - T_i \cos\theta \sinh 2s)$, which is minimum for a squeezing angle $\theta = 2 \pi m$, $m \in \mathbb{Z}$ indicating that displacement and squeezing should be along the same direction.

Using error propagation, we can readily obtain the uncertainty in estimating $C$ given in Eq. (\AB{9}) in the main text
\begin{align}
    \Delta^2 C &= \abs{\frac{\partial C}{\partial R}}^{2} \Delta^2 R
    = \abs{\frac{\partial R}{\partial C}}^{-2} R^2 \left(\frac{\Delta^2 \hat{n}_R}{\langle \hat{n}_R \rangle^2} + \frac{\Delta^2 \hat{n}_L}{\langle \hat{n}_L \rangle^2} \right)
    = \frac{1}{(\abs{\alpha} \Delta \epsilon\,l\, \ln 10)^2} \left[2(e^{-2s}-1) + 1/T_R + 1/T_L \right]\,.
\end{align}
\end{document}